\documentclass[12pt]{iopart}
\usepackage{dsfont}
\usepackage{amssymb}
\usepackage{mathbbol,bbm}
\usepackage{graphicx,float}
\usepackage[final]{showkeys}

\usepackage{cite}
\usepackage{iopams}
\begin{document}

\title{Sub-Heisenberg estimation of non-random phase-shifts}
\author{\'{A}ngel Rivas$^{1}$ and Alfredo Luis$^{2}$}

\address{$^1$ Departamento de F\'{\i}sica Te\'orica I,
Facultad de Ciencias F\'{\i}sicas, Universidad Complutense,
28040 Madrid, Spain \\
$^2$ Departamento de \'{O}ptica, Facultad de Ciencias
F\'{\i}sicas, Universidad Complutense, 28040 Madrid, Spain
}

\ead{alluis@fis.ucm.es}
\begin{abstract}
We provide evidence that the uncertainty in detection of small and deterministic phase-shift deviations from a working point can be lower than the Heisenberg bound, for fixed finite mean number of photons. We achieve that by exploiting non-linearity of estimators and coherence with the vacuum.
\end{abstract}

\pacs{42.50.St, 03.65.-w, 42.50.Dv}
\maketitle

\section{Introduction}
Metrology is a prime issue both from theoretical and practical reasons.
Precise measurements are crucial in physics since they constitute
the link between the theory and nature, so that accurate measurements can
promote or reject a theory.

The key contribution of the quantum theory to metrology is that quantum
fluctuations limit the resolution. To be more precise, quantum limits
emerge when we impose constraints. The most popular restriction is to
consider a fixed finite mean number of particles, although time limitations
might also be considered \cite{Heisenberg1}.

The existence of quantum metrology limits imposed by finite resources
is very clear if we consider systems with bounded number of particles
\cite{Heisenberg1,Heisenberg2,Heisenberg3}. In linear schemes this is the Heisenberg
bound that states that the uncertainty is lower bounded by the inverse of the total
number of particles employed in the measurement. However, most states
have unlimited number of particles even if their mean number is finite.
In such a case, despite that it may be expected a similar scaling law with the total number of particles as in the finite case,
to give a ultimate bound to the sensitivity turns out to be considerably more intricate
\cite{SSW,BLC,HS,limitnvariable2,limitnvariable3,Augusto,Dowling,Joo,GLM,Hall,Tsang,Nair,Gao}.

In this work we provide evidence that the Heisenberg bound can be beaten
with a fixed and finite mean number of particles. Sub-Heisenberg resolution arises because of the combination
of two effects. Firstly, performance estimators (such as Fisher information
\cite{FI,BC}) may be nonlinear functions of the photon number, even if we are
dealing exclusively with linear processes. Secondly, the use of probes in states including coherent superpositions with the
vacuum shifts the number distribution to larger photon numbers but keeps fixed the mean number.

\section{Quantum limits in the detection of non-random signals}
Throughout the article we consider that the signal is encoded as a shift in the phase of a quantum harmonic oscillator; a suitable practical implementation may be a single-mode electromagnetic field. Furthermore we consider non-random signals, i.e. the phase-shift is assumed to be a unknown but deterministic signal \cite{Levy,Poor,Casella} which is not subject to statistical fluctuations. Rather, it is the estimation process itself what is the source of randomness because the statistical nature of the quantum mechanics. For instance, parameters such as the mass of a particle or the amplitude of a gravitational wave (on a classical theory of gravitation) are commonly assumed to be deterministic signals.

As a measurement of uncertainty we shall consider the mean squared error,
\begin{equation}\label{MSE}
(\Delta\tilde{\phi})^2=\int_I dx p(x|\phi)[\tilde{\phi}(x)-\langle \tilde{\phi}\rangle]^2,
\end{equation}
here $x$ denotes the possible outcomes of the performed measurement which takes value in the measurable set $I$.
The outcomes $x$ follow the distribution $p(x|\phi)$ that depends on the true value of the phase-shift
$\phi$. The function $\tilde{\phi}(x)$ is an estimator of the phase-shift with expectation value
$\langle \tilde{\phi} \rangle = \int_I dx p(x|\phi) \tilde{\phi}(x)$. For an unbiased estimator $\langle
\tilde{\phi}\rangle=\phi$.

Note that the sensitivity of a measurement may depend on the unknown value of the signal $\Delta\tilde{\phi}\equiv(\Delta\tilde{\phi})_\phi$. In fact, this is quite natural in real experiments (see \cite{sdu} for example). Since quantum fluctuations are typically relevant only for small signals, we shall focus here on the behaviour of $(\Delta\tilde{\phi})_\phi$ for small phase shifts $\phi\ll1$. Moreover we are primarily concerned with the resolution for a specific value of the nonrandom signal, say $\phi=0$. This might be the case of the measurement of the photon mass $M_{\rm ph}$ so that $\phi \propto M_{\rm ph}$ \cite{mpm}. Accordingly we care mainly for the uncertainty $(\Delta\tilde{\phi})_{\phi=0}$ at that specific point $\phi=0$ disregarding its behavior at other signal values.

In this regard, it is important to note the difference between $(\Delta\tilde{\phi})_\phi$ and the averaged mean squared error, which is typical in Bayesian estimation approaches. In those situations, the signal is also considered to be a random parameter, and the uncertainty may be measured by the formula
\begin{equation}\label{MSEB}
\fl \delta\tilde{\phi}^2=\int_{-\pi}^{\pi}d\phi \int_I  dx p(\phi) p(x|\phi)[\tilde{\phi}(x)-\phi]^2=\int_{-\pi}^{\pi}d\phi \int_I  dx p(x,\phi)[\tilde{\phi}(x)-\phi]^2,
\end{equation}
where $p(\phi)$ is the so-called prior probability. It may represent somehow the state of ignorance about $\phi$ prior to the experiment \cite{Casella}. Note that for an unbiased estimator Eq. (\ref{MSEB}) becomes
\begin{equation}
\delta\tilde{\phi}^2=\int_{-\pi}^{\pi}d\phi \int_I p(x,\phi)[\tilde{\phi}(x)-\langle \tilde{\phi}\rangle]^2=\langle(\Delta\tilde{\phi})^2\rangle,
\end{equation}
which coincides with the mean squared error (\ref{MSE}) averaged with the prior probability.

In recent papers \cite{GLM,Hall,Tsang,Nair,Gao} several bounds have been established for the averaged error
$\delta\tilde{\phi}$ which behave like $\kappa/\bar{N}_{\rm T}$. Here $\kappa$ is a constant of the order of one
and $\bar{N}_{\rm T}$ is the mean value of the total number of photons employed in the estimating procedure.

However, as already mentioned, in this work we are interested in the non-averaged uncertainty $\Delta \tilde{\phi}$. Previously
\cite{limitnvariable2} suggested that $\Delta \tilde{\phi}$ is always larger or of the same
order as $1/\bar{N}_{\rm T}$. In \cite{Augusto} it was proved that, within a two-field-mode context, $\Delta
\tilde{\phi}$ is lower bounded exactly by $1/\bar{N}_{\rm T}$ for states and measurements without
coherences between different subspaces of fixed total photon number. That, in one-mode approximation, corresponds to $1/(2\bar{N}_{\rm T})$. We shall refer to $1/(2\bar{N}_{\rm T})$ as the ``Heisenberg bound''. For the general case \cite{Augusto} only argues that the Heisenberg bound strictly applies in the asymptotic limit when the number of repetitions becomes infinity. So $\Delta \tilde{\phi}$ could still be lower than
$1/(2\bar{N}_{\rm T})$ for finite number of measurements, as explicitly recognized in \cite{Augusto}. This is examined
in Sections 3 and 4.

For unbiased estimators, a very well-known inequality is the Cram\'er-Rao bound \cite{FI,BC}, which is formulated for $(\Delta\tilde{\phi})^2$,
\begin{equation}\label{CramerRao}
(\Delta\tilde{\phi})^2\geq\frac{1}{mF(\phi)}\geq\frac{1}{mF_Q(\phi)},
\end{equation}
where $F(\phi)$ is the Fisher information of the measurement, $m$ the number of repetitions of it, and $F_Q(\phi)$ the quantum Fisher information which corresponds to the best possible choice of the measurement \cite{BC}. If such a measurement is good enough, this bound can be achieved in the asymptotic limit $m\rightarrow\infty$ by using the maximum likelihood estimator \cite{Levy,Poor,Casella}.

For linear schemes, $F_Q(\phi)$ is of the form $\alpha \bar{n}^2 + \beta \bar{n}$, where usually $\alpha$, $\beta$ do not depend on the probe state. Hence (\ref{CramerRao}) asserts that the optimum mean squared error must decrease as $1/\bar{n}$, which is in accordance to the result of \cite{Augusto}. Thus, we refer to $\Delta\tilde{\phi} \propto 1/\bar{n}$ (i.e. $\Delta\tilde{\phi} \propto 1/ \bar{N}_{\rm T}$) as ``Heisenberg scaling''. Sensitivities below the Heisenberg bound $1/(2\bar{N}_{\rm T})$ may be compatible with the Heisenberg scaling, but the implication in the reverse sense is not true, so that if the Heisenberg scaling is beaten then the Heisenberg bound is certainly surpassed at least for some $\bar{N}_{\rm T}$ large enough.

In principle, to break the Heisenberg scaling we may resort to non-linear schemes \cite{Heisenberg3,AlfredoPLA} where $F_Q(\phi)$ is polynomial of higher order in $\bar{n}$. However, it might be broken also in linear schemes if $\alpha$ and/or $\beta$ in $F_Q\sim\alpha \bar{n}^2 + \beta
\bar{n}$ depend on the mean number $\bar{n}$ of the probe state. We shall examine this possibility in Section 4.

\section{Probe states and signal transformation}

The structure of any signal-detection process, classical or quantum, is quite universal. A probe experiences a signal-dependent transformation. The change of the probe state is monitored by a measurement whose outputs serve to infer the value of the signal.

Next we examine the construction of a probe state with the aim to beat the Heisenberg bound $1/(2\bar{N}_{\rm T})$.

\subsection{Squeezed states}

As a first example let us consider the familiar choice \cite{CavesComp} of a quadrature squeezed state
\cite{SZ},
\begin{equation}
| \xi \rangle = D(\bar{y}) S(r ) | 0 \rangle ,
\end{equation}
with
\begin{equation}
D(\bar{y} ) = \exp \left ( i \bar{y} X /2 \right ) ,
\qquad S(r) = \exp \left [ i r \left ( X Y + Y X \right )/4 \right ] ,
\end{equation}
where $X$, $Y$ are the quadratures of a single-mode field of complex amplitude $a$,
\begin{equation}
X = a^\dagger + a, \qquad
Y = i \left (  a^\dagger - a \right ) ,
\end{equation}
with $[X,Y] = 2 i$. Moreover we have
\begin{equation}
\langle X \rangle_\xi = 0 , \quad
\langle Y \rangle_\xi = \bar{y} ,
\end{equation}
with
\begin{equation}
\bar{n}_\xi = \langle a^\dagger a \rangle_\xi =
\frac{\bar{y}^2}{4} + \sinh^2 r, \qquad
\left ( \Delta X \right )_\xi^2 = \exp (-2r ) .
\end{equation}
For definiteness, throughout we will assume $\bar{y} \gg 1$ and $r \gg 1$. In addition we
take a precisely equal splitting of the photons between the coherent and squeezed parts,
$\bar{y}^2/4=\sinh^2 r$. Despite this is not the best distribution of the energy resources
\cite{Monras}, it does not  qualitatively affect the results and simplifies subsequent
computations. The following relations hold approximately
\begin{equation}
\label{rel}
\bar{y}^2 \simeq e^{2r} \simeq \frac{1}{\left ( \Delta X \right )_\xi^2}
\simeq 2 \bar{n}_\xi \gg 1.
\end{equation}

As usual, in linear schemes the phase shift to be detected $\phi$ is generated by the number operator $a^\dagger a$ so that the transformed
probe state is $\exp (-i \phi a^\dagger a ) | \xi \rangle$. In this situation the quantum Fisher information is four times the variance of $a^\dagger a$ in the state $| \xi \rangle$ \cite{BC}, and yields
\begin{equation}
F_{Q, \xi}=4 \left ( \Delta a^\dagger a \right )^2\simeq 6 \bar{n}^2.
\end{equation}

In order to discuss  whether the above result for the quantum Fisher information allows for an improvement on the Heisenberg bound, it may be convenient to consider a practical scheme, even if this is not fully optimal. To this end let us analyze the homodyne detection measuring the quadrature $X$ in the transformed probe state $\exp (-i \phi a^\dagger a ) | \xi \rangle$.
In this case the Cram\'{e}r-Rao lower bound reads \cite{FI}
\begin{equation}
\left ( \Delta \tilde{\phi} \right )^2_{\phi} \geq \frac{1}{m F(\phi)} ,
\end{equation}
where $m$ is the number of repetitions of the measurement and the Fisher information reads
\begin{equation}
\label{FI}
F = \int dx \frac{1}{p(x|\phi)} \left [ \frac{d p(x|\phi)}{d \phi}
\right ]^2.
\end{equation}

For a small enough signal $\phi \ll 1$ we may approximate
\begin{equation}\label{xitrans}
\langle x | e^{-i\phi a^\dagger a} | \xi  \rangle \simeq
\frac{1}{(2 \pi)^{1/4} \sqrt{(\Delta X)_\xi}} \exp \left \{ i\left[
\frac{\bar{y} x}{2}-g(x)\phi\right] - \frac{[x-\bar{x}(\phi)]^2}{4 (\Delta X )_\xi^2}
\right \} ,
\end{equation}
where
\begin{equation}\label{gx}
g(x)=\frac{1}{4} \left ( x^2 - 2 +
\frac{2}{(\Delta X)_\xi^2} + \bar{y}^2 - \frac{x^2}{(\Delta X)_\xi^4}\right),
\end{equation}
and
\begin{equation}
\bar{x} (\phi) = \bar{y} \sin \phi \simeq \bar{y} \phi.
\end{equation}
Note that the change in $\langle Y(\phi)\rangle=\bar{y}+\mathcal{O}(\phi^2)$ and $\Delta X(\phi)=\Delta X+\mathcal{O}(\phi^2)$ is of second order in $\phi$, because $\langle X \rangle = 0$ and $\langle XY+YX \rangle= 0 $.

Under this assumption the statistics of the measurement is a Gaussian function centered in $\bar{x} (\phi)=\bar{y} \phi$,
\begin{equation}
p_\xi(x|\phi) = |\langle x | e^{-i\phi a^\dagger a} | \xi  \rangle|^2=\frac{1}{\sqrt{2 \pi} (\Delta X)_\xi } \exp \left \{ -
\frac{[x-\bar{x}(\phi)]^2}{2 (\Delta X )_\xi^2} \right \}  ,
\end{equation}
leading to
\begin{equation}
\label{ssp}
F_\xi  = \frac{\bar{y}^2}{( \Delta X )_\xi^2} \simeq 4 \bar{n}^2 ,
\qquad
\left ( \Delta \tilde{\phi} \right )_{\xi,\phi} ^2 \geq \frac{1}{4 m \bar{n}^2}.
\end{equation}

Since the estimation of $\phi$ is equivalent to the estimation of the mean of a Gaussian distribution, the maximum likelihood estimator reaches the above Cram\'er-Rao bound for any $m$ \cite{Levy,Poor,Casella}. Thus, the best that can be done with a squeezed probe is the single-trial sensitivity, $m=1$, $\bar{N}_{\rm T}=m\bar{n}=\bar{n}$ so that $\left ( \Delta \tilde{\phi} \right )_{\xi,\phi}  \simeq 1/(2 \bar{N}_{\rm T})$, which is the Heisenberg bound. Therefore, to obtain sub-Heisenberg resolution, we must resort to another state.

\subsection{Superpositions of vacuum and Squeezed states}

Instead of a squeezed state, consider the probe in the superposition
\begin{equation}
\label{probe}
| \psi \rangle = \mu | 0 \rangle + \nu | \xi \rangle ,
\end{equation}
of the vacuum $| 0 \rangle$ and a squeezed coherent state $| \xi \rangle$ in the same terms as in the previous section. The parameters $\mu, \nu$ are assumed to be real for simplicity. Under the conditions (\ref{rel}) the states $| 0 \rangle$ and $| \xi
\rangle$ are approximately orthogonal
\begin{equation}
\langle 0 | \xi \rangle = \sqrt{2 (\Delta X)_\xi} \exp \left [- \bar{y}^2
(\Delta X)_\xi^2 /4 \right ] \propto \frac{1}{\bar{n}_\xi^{1/4}} \ll 1 ,
\end{equation}
so that $\mu^2 + \nu^2 \simeq 1$. Furthermore, we shall consider
$\nu \ll 1$, $\mu \simeq 1$.

The mean number of photons of the state (\ref{probe}) reads
\begin{equation}
\bar{n} = \langle \psi | a^\dagger a | \psi \rangle =
\nu^2 \bar{n}_\xi .
\end{equation}
Since our purpose is to investigate quantum limits for fixed
finite resources we will suppose that $\bar{n}$ is fixed, so
that the number of photons in the squeezed state $| \xi \rangle$
depends on $\nu$ as
\begin{equation}\label{nxi}
\bar{n}_\xi = \frac{\bar{n}}{\nu^2}.
\end{equation}

In this case, considering $\bar{n}_\xi \gg1 $ and $\nu \ll 1$
\begin{equation}
\left ( \Delta a^\dagger a \right )^2 \simeq \frac{5}{2} \nu^2
\bar{n}_\xi^2 - \nu^4 \bar{n}_\xi^2 \simeq \frac{5}{2}
\frac{\bar{n}^2}{\nu^2} ,
\end{equation}
and
\begin{equation}
\label{qFi}
F_Q \simeq \frac{10 \bar{n}^2}{\nu^2},
\qquad (\Delta \tilde{\phi} )^2 \geq \frac{\nu^2}{10 m \bar{n}^2}.
\end{equation}
These results suggest that $|\psi\rangle$ is able to estimate phase changes with an uncertainty far
below the Heisenberg bound, as we have taken $\nu \ll 1$. However, in order to confirm this, we need to find (if they exist) a
good enough measurement scheme and estimator function.

Let us consider again the homodyne detection measuring the quadrature $X$ in the transformed probe state $\exp (-i \phi a^\dagger a ) | \psi \rangle$. Up to first order in $\phi$ and $\nu$ the statistics yields
\begin{equation}\label{p(x)}
\fl p(x|\phi)\simeq\frac{1}{\sqrt{2 \pi}}\left(\mu^2 \exp \left(-\frac{x^2}{2} \right) +  \frac{2\mu\nu}{
\sqrt{( \Delta X)_\xi}} \cos\left[\frac{\bar{y} x}{2}-\phi g(x)\right]\exp\left\{-\frac{x^2}{4}-\frac{(x-\bar{y}\phi)^2}{4 (\Delta X)_\xi^2 }\right\}\right),
\end{equation}
where $g(x)$ is given by (\ref{gx}).

This approximation can be checked by expanding $p(x|\phi)$ in power series of $\phi$ and comparing with the result of $|\langle x|\exp (-i \phi a^\dagger a ) | \psi \rangle|^2\simeq|\langle x|1 -i \phi a^\dagger a  | \psi \rangle|^2$ at first order in $\nu$. Here for the vacuum we have
\begin{equation}
\langle x |e^{-i\phi a^\dagger a} |0 \rangle=\langle x  |0 \rangle = \frac{1}{(2 \pi)^{1/4}} \exp (-x^2 /4 ),
\end{equation}
while Eq. (\ref{xitrans}) holds for the squeezed component.

For the computation of the Fisher information, in the denominator in Eq. (\ref{FI})
we can safely approximate $p(x|\phi) \simeq \langle x | 0 \rangle^2 $. Taking into
account relations (\ref{rel}), i.e., $\bar{y} \simeq 1/ ( \Delta X )_\xi \simeq
\sqrt{2 \bar{n}_\xi} \gg 1$, we get that, after a long but straightforward calculation,
the leading term in $\bar{n}_\xi$ is
\begin{equation}
\label{Fcs}
F \simeq 4 \nu^2 \bar{n}_\xi^2 \simeq 4 \frac{\bar{n}^2}{\nu^2} ,
\end{equation}
so that
\begin{equation}
\label{res}
\left ( \Delta \tilde{\phi} \right )^2 \geq \frac{\nu^2}{4 m \bar{n}^2} .
\end{equation}
This scales as the quantum Fisher information (\ref{qFi}), so that the
conclusion is the same as above.
\begin{figure}
\begin{center}
\includegraphics[width=0.9\textwidth]{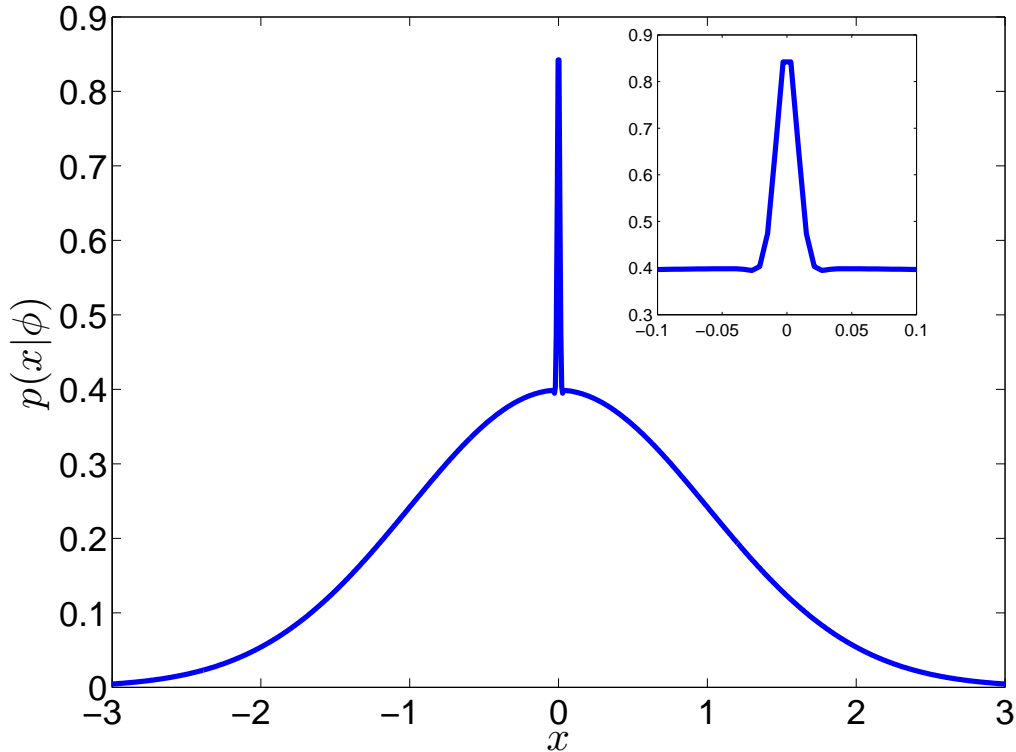}
\end{center}
\caption{Probability distribution of the quadrature $X$ in the
probe state (\ref{probe}) as a function of $x$ for $\bar{n}=25$
and $\nu =0.05$. The inset represents in more detail the central
peak.}
\end{figure}

For the sake of clarity, in Fig. 1 we have plotted $p(x|\phi)$ for $\bar{n}=25$, $\phi = 0$, and $\nu =5\times10^{-2}$. It can be appreciated that the squeezed state contributes with a very narrow central Gaussian peak in a more uniform background provided by the vacuum. For $\phi$ small the effect of the phase shift becomes a displacement of the small peak proportional to $\phi$. This peak is the cause of the increased Fisher information depending on $\nu$. However note that the Fisher information is only a lower bound for the resolution, which is only attainable for Gaussian distributions. Since the complete distribution (\ref{p(x)}) is manifestly nonGaussian, the lower bound cannot be achieved. Hence, the performance of the bound can only be tested by actual calculation with a specific estimator. That will be the aim of the next section.

On the other hand, it is interesting to estimate how likely is the outcome $x$ to fall below the tiny peak. Since the probability at $x=0$ is $p(x=0)\simeq \left(\mu^2+2\mu\nu/\sqrt{(\Delta X)_{\xi}}\right)/\sqrt{2 \pi}$, while its width is proportional to $(\Delta X)_{\xi}$, we
obtain the probability below the small peak to be $p(x=0)(\Delta X)_{\xi} \simeq \nu/\sqrt{\bar{n}}+\nu^{3/2}/\bar{n}^{1/4}\simeq\nu/\sqrt{\bar{n}}$,
where this approximation holds for $\nu\ll1$.

\section{Results and Discussion}

In order to check that the state (\ref{probe}) can actually beat the Heisenberg bound we have to choose an appropriate estimator. To that aim we consider the maximum likelihood estimator, $\tilde{\phi}=\tilde{\phi}_{\rm ML}$, such that
\begin{equation}\label{est}
\mathcal{L}(\tilde{\phi}_{\rm ML}|x_1,x_2,\ldots,x_m)=\max_{\phi}\mathcal{L}(\phi|x_1,x_2,\ldots,x_m),
\end{equation}
where the likelihood function is
\begin{equation}
\mathcal{L}(\phi|x_1,x_2,\ldots,x_m)=\prod_{i=1}^m p(x_i|\phi).
\end{equation}
The maximum likelihood estimator is unbiased for any value of $\phi$ and it is asymptotically efficient \cite{Levy,Poor,Casella}, i.e. it reaches the Cram\'er-Rao bound for a large number of measurements $m\rightarrow\infty$.

However for large $m$, Monte Carlo simulations seem to indicate that the sensitivity is worse than the Heisenberg bound. That basically happens because the uncertainty decreases with $m$ as $1/\sqrt{m}$ whereas the Heisenberg bound does it as $1/m$. Thus we restrict our study to the best situation possible where $m=1$. The table 1 shows some of the uncertainties obtained by Monte Carlo simulations for different mean number of photons.

\begin{table}[h]
\begin{center}
\begin{tabular}{|c|c|c|c|}
\hline
$\bar{n}$ & $\tilde{\phi}$ & $(\Delta\tilde{\phi})_{\phi=0}$ & $1/(2\bar{N}_{\rm T})$ \\
\hline
1 & $1.60\times10^{-4}$ & $0.035$ & $0.500$ \\
\hline
2 & $-9.50\times10^{-5}$ & $0.025$  & $0.250$\\
\hline
3 & $8.79\times10^{-5}$ & $0.020$ & $0.167$ \\
\hline
4 & $-5.43\times10^{-5}$ & $0.017$ & $0.125$\\
\hline
5 & $4.91\times10^{-5}$ & $0.016$ & $0.100$ \\
\hline
\end{tabular}
\caption{Results of the Monte Carlo simulation for $\phi=0$ and different mean number of photons. It has been taken $\nu=0.05$ and $m=1$. The two last columns show clear violations of the Heisenberg bound $1/(2\bar{N}_{\rm T})$.}
\end{center}
\end{table}

As it can be seen, the probe state (\ref{probe}) provides sensitivities considerably smaller than the Heisenberg bound, but compatible with the Cram\'er-Rao bound (\ref{res}). For instance, for $m=\bar{n}=1$, $\nu =0.05$, it reads $(\Delta\tilde{\phi})_{\phi=0} \geq 0.025$. On the other hand, in Fig. 2 we have plotted the evolution of the uncertainty $(\Delta\tilde{\phi})_{\phi=0}$ with the number of points taken to simulate the experiment with $\bar{n}=1$ by the Monte Carlo procedure. It clearly shows that statistical uncertainties are small enough to reject them as the possible reason for the violation of the Heisenberg bound.

\begin{figure}
\begin{center}
\includegraphics[width=0.9\textwidth]{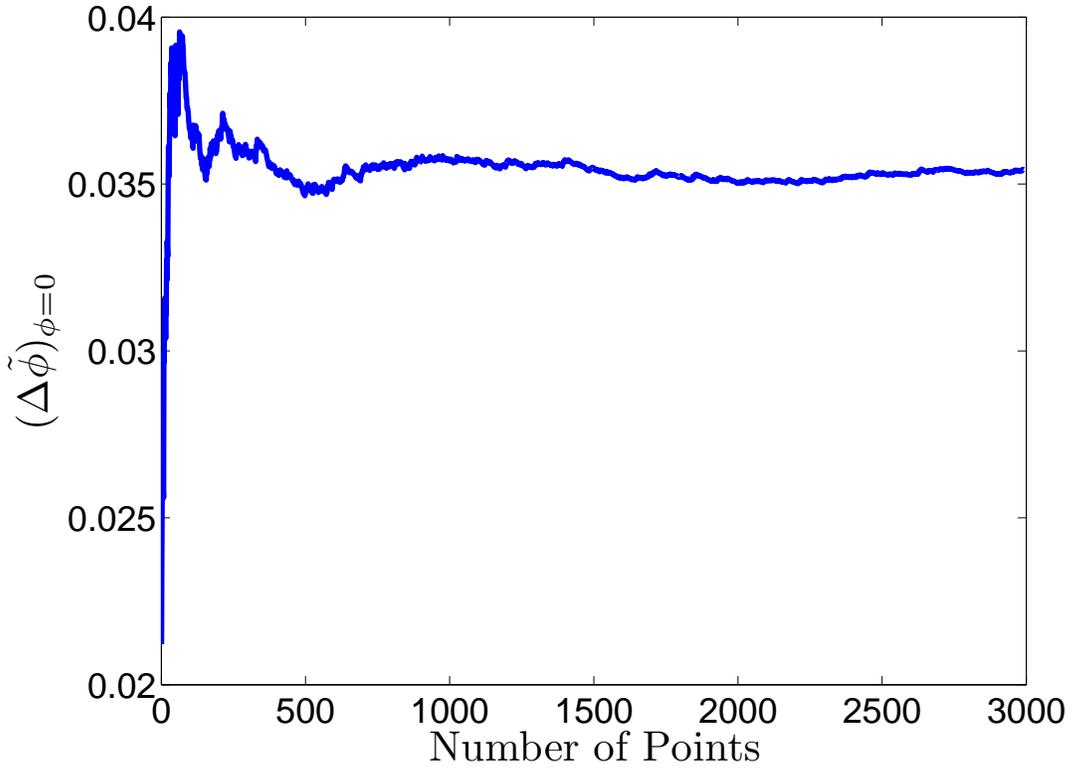}
\end{center}
\caption{Evolution of the uncertainty $(\Delta\tilde{\phi})_{\phi=0}$ with the number of points of the Monte Carlo simulation for $m=1$, $\nu=0.05$ and $\bar{n}=1$.}
\end{figure}

Similar uncertainties are obtained up to the fourth decimal place for any value of $\phi$ inside of $(\Delta\tilde{\phi})_{\phi=0}$. Of course, for the range of parameters we have considered, it is not possible to distinguish between such a finite $\phi$ and $\phi=0$. However this state is just intended to be an example of ``proof of principle'' that the Heisenberg bound may be suppressed.

Interestingly, we may also explore whether the probe state (\ref{probe}) can beat the Heisenberg scaling.
At difference with previous approaches, in our case the uncertainty depends on an extra free parameter
$\nu$ that depends on the probe state. Thus, we are free to consider that $\nu$ might depend on $\bar{n}$ for some probe states.
Note that provided the mean number of squeezed photons is given by Eq. (\ref{nxi}), $\bar{n}_\xi = \bar{n}/\nu(\bar{n})^2$,
the total number of photons $\bar{n}$ is not affected by $\nu(\bar{n})$ despite of its dependence with $\bar{n}$. Fig. 3 shows
the results of a Monte Carlo simulation for $\nu=0.05/\bar{n}$. The linear fitting of the data provides a scaling
law of the form $(\Delta\tilde{\phi})_{\phi=0}\sim1/\bar{N}_{\rm T}^{1.4976\pm0.0098}$, which improves the Heisenberg scaling
$(\Delta\tilde{\phi})_{\phi=0}\sim1/\bar{N}_{\rm T}$.

\begin{figure}
\begin{center}
\includegraphics[width=0.9\textwidth]{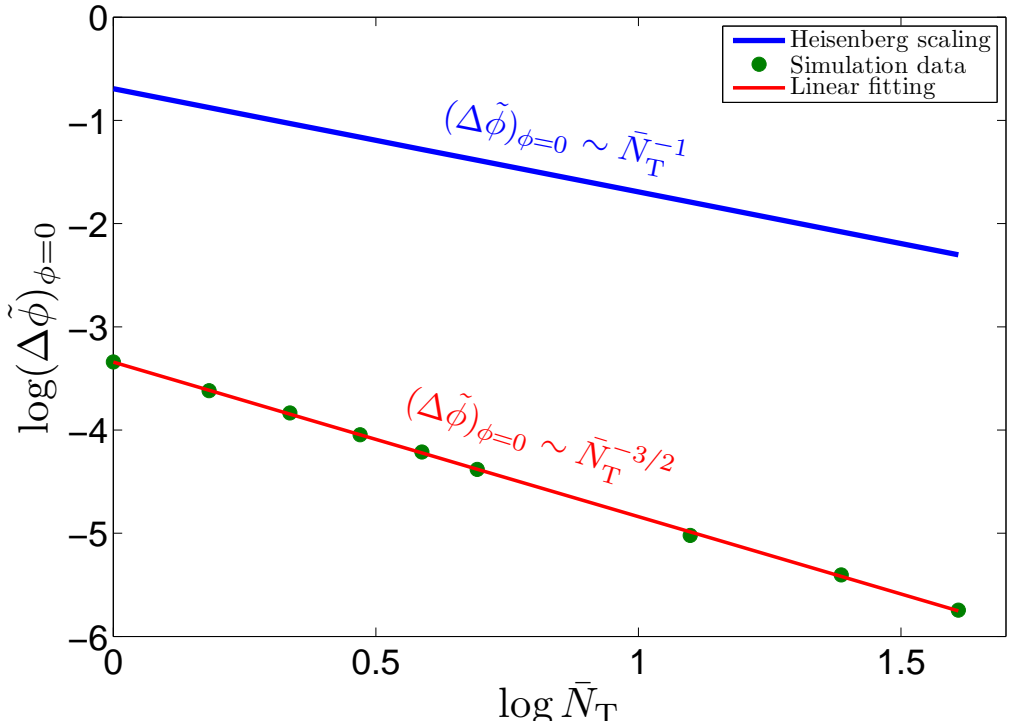}
\end{center}
\caption{Logarithmic plot of the sensitivity $(\Delta\tilde{\phi})_{\phi=0}$ with $m=1$ and $\nu=0.05/\bar{n}$ as a function of $\bar{n}$, for some values between 1 and 5. The linear fitting (red line) leads to a sub-Heisenberg scaling $(\Delta\tilde{\phi})_{\phi=0}=(0.0354\pm0.0003)/\bar{N}_{\rm T}^{1.4976\pm0.0098}$.}
\end{figure}

It is remarkable that in \cite{SSW} another scheme which seemed to beat the Heisenberg bound was proposed. Incidentally our Fig. 1 presents some similarities with the figure in \cite{SSW}. However, in contrast with our case, Shapiro et al. \cite{SSW} focus on the results of a covariant measurement, where the uncertainty is the same for any value of $\phi$. For that case, Braunstein et al. \cite{BLC} showed that the sensitivity neither improves the Heisenberg bound nor the Heisenberg scaling. This is actually consistent with the results in \cite{GLM,Hall,Tsang,Nair,Gao}, where the averaged mean squared error (\ref{MSEB}) is analysed. Our approach is different from \cite{SSW} and \cite{BLC} in the sense that we do not look for a $\phi-$independent sensitivity, but a large sensitivity at a very small interval around one point (namely $\phi=0$). Because of this we can resort to homodyne measurements which are not covariant.

Thus, we want to emphasize that the improved resolution holds only for $\phi$ close enough to $\phi=0$. Nevertheless, the limitation of improved resolution to small intervals of signal values is quite frequent in practice. For example in \cite{sdu} improved sensitivity holds just for a signal interval $\delta \phi$ of the same order of the uncertainty $\delta \phi \simeq \Delta \phi$. Therefore, it appears that the condition $\Delta \phi \ll \delta \phi$ is not mandatory in real experiments.

\section{Conclusion}

Quantum metrology is framed by some implicit understandings that may limit its development. In this work we have shown that
sub-Heisenberg resolution is possible for local estimators. Concretely we give an example where homodyne measurement provides uncertainties smaller than $1/(2\bar{N}_{\rm T})$ for signals $\phi$ close enough to $\phi=0$. This has to be contrasted with previous results  involving covariant measurements schemes \cite{SSW,BLC}, asymptotic number of repetitions \cite{Augusto}, or the recent rigorous proof of sensitivity bounds for averaged mean squared errors in Bayesian estimation \cite{GLM,Hall,Tsang,Nair,Gao}.

The reason for this improvement lies in the nonlinear behavior of performance estimators with the number of photons. This leads to
resolution improvement at fixed mean number of photons because of the shift in the photon-number statistics caused by the
coherent superposition with the vacuum.

Local measurement and uncertainties may be especially useful if the experimenter knows a priori that the signal is very close to a certain value. However our aim here has been just to provide evidence that resolutions beyond the Heisenberg bound are possible. We hope that may disclose a new perspective on quantum metrology avoiding previously assumed performance limits.

\ack

We are grateful to Profs. A. Smerzi, L. Maccone, M. Tsang and M. J. W. Hall
for enlightening discussions. We thank financial support from project
QUITEMAD S2009-ESP-1594 of the Consejer\'{\i}a de Educaci\'{o}n
de la Comunidad de Madrid. A. R. acknowledges MICINN FIS2009-10061.
A. L. acknowledges support from project No. FIS2008-01267 of the
Spanish Direcci\'{o}n General de Investigaci\'{o}n del Ministerio
de Ciencia e Innovaci\'{o}n.

\end{document}